\documentclass[11pt]{article}

\usepackage[margin=25mm]{geometry}

\usepackage{mathptmx, parskip, nomencl, framed, longtable, setspace, adjustbox,multicol}
\usepackage{subcaption}
\usepackage{float,graphicx, booktabs, tabularx}
\usepackage{setspace}
\usepackage[numbers]{natbib}
\renewcommand{\nomgroup}[1]{
\ifthenelse{\equal{1}{G}}{\item[\textbf{Greek symbols}]}
\ifthenelse{\equal{1}{P}}{\item[\textbf{Parameters}]}
\ifthenelse{\equal{1}{S}}{\item[\textbf{Subscripts}]}
\ifthenelse{\equal{1}{X}}{\item[\textbf{Superscripts}]}
\ifthenelse{\equal{1}{Z}}{\item[\textbf{Abbreviations}]}
}

\makenomenclature
\setcitestyle{maxcitenames=2}
\begin{document}

\begin{center}
\textbf{\MakeUppercase{{\fontsize{14pt}{14pt}\selectfont Start-up and Transient Response Characteristics of Natural Draft Direct Dry Cooling Systems at Various Scales}}}
\end{center}
\vspace{18pt}

\begin{center}
{\fontsize{10pt}{10pt}\selectfont
Wian Strydom\textsuperscript{1,*}, Johannes Pretorius\textsuperscript{1}, Ryno Laubscher\textsuperscript{1}
}
\end{center}
\begin{center}
{\fontsize{10pt}{10pt}\selectfont
\textsuperscript{1}Department of Mechanical and Mechatronic Engineering, University of Stellenbosch,\textsuperscript{*}wian.str.98@gmail.com \\
}
\end{center}


\section*{{\fontsize{11pt}{0pt}\selectfont Abstract}}


Natural draft direct dry cooling systems (NDDDCSs) combine the benefits of two traditional dry cooling systems, air-cooled condensers (ACCs) and indirect natural draft dry cooling systems, to offer advantages such as reduced system complexity, high thermal efficiency, and low parasitic power losses and operational costs. These scalable systems can potentially operate in both large coal-fired power plants (900~MWt) and medium-scale concentrated solar power (CSP) plants (100~MWt). While prior research focused on NDDDCS steady-state performance, this study delves into the transient behaviour and start-up characteristics of this cooling system, crucial for power plants with rapid start-up requirements or load ramps during electricity demand peaks. The paper employs a transient 1-D mathematical model to characterize the NDDDCSs' response to start-up and load ramp conditions under no-wind, presenting results for both large- and medium scales while identifying limiting factors. 


{\fontsize{11pt}{0pt}\selectfont \noindent\textbf{Keywords:} Natural draft, transient, start-up, condenser, dry cooling}
\section*{{\fontsize{12pt}{0pt}\selectfont 1. Introduction}}

Climate change challenges the energy generation industry, through increasing water scarcity \cite{Waterscarce}, energy demand, and global population growth \cite{council2013world}. These factors necessitate a shift towards more resource-efficient energy generation technologies. Dry cooling technologies, with reduced water usage compared to wet cooled systems \cite{osti_1009674}, are natural candidates.

Two prevalent dry cooling systems, forced draft air-cooled condensers (ACCs) and indirect natural draft dry cooling systems, offer distinct advantages. The ACCs operate using a multitude of large fan drives and have a high thermal efficiency. However, the fans run virtually continuously in harsh conditions and require regular maintenance. These systems also have significant parasitic losses. Indirect natural draft systems, while avoiding fan drives, face challenges in reduced thermal efficiency and high complexity which demands substantial initial capital investments \cite{CONRADIE199825}.

The natural draft direct dry cooling system (NDDDCS) combines the strengths of ACCs and indirect systems, offering higher thermal efficiencies and lower operational costs. The absence of fan drives eliminates several failure modes, as passive airflow is driven by a natural draft. Limited research has been performed on NDDDCSs to date, and have focused on steady-state performance. Studies by Kong et al. using 3-D computational fluid dynamics (CFD) simulations highlight the superiority of vertically arranged heat exchanger bundles, with increased performance associated with larger apex angles \cite{KongYanqiang2018Tpon, KongYanqiang2018Aaac}. Normal and radiator type windbreaker walls further enhance performance under crosswind conditions \cite{KongYanqiang2018Wlti, goodarzi2013heat}.

Following from Kr$\mathrm{\ddot{o}}$ger's work on ACCs and indirect systems \cite{alma990005992320803436}, 1-D modelling techniques have proven viable for NDDDCS investigations. Strydom et al.'s sensitivity analysis emphasized the impact of geometric variation on NDDDCS performance, while also showcasing its potential as a standalone cooling system for CSP plants \cite{strydom_HEFAT, strydom_ENFHT}.

This study addresses a gap in the literature by exploring the start-up characteristics of NDDDCSs, crucial for power generation cycles during rapid start-ups or demand peaks. Transient performance is gaining significance as intermittent renewable technologies become more prevalent on the grid. Existing transient investigations have focussed on small-scale indirect natural draft dry cooling towers \cite{DONGstart,DONGweather,DONGwind}, leaving a knowledge gap concerning NDDDCS start-up. The NDDDCS relies on the interdependence of the condensation process and system draft, unlike indirect systems which mechanically pump water into the heat exchangers.

This research develops transient 1-D mathematical models to quantify NDDDCS start-up and transient load ramp characteristics at two scales: large-scale coal-fired power plant (900~MWt) and medium-scale CSP plant (100~MWt). Results shed light on NDDDCS response time and potential bottlenecks in start-up and load ramp scenarios.


\section*{{\fontsize{12pt}{12pt}\selectfont 2. Reference Tower Geometry and Sizing}}

\begin{figure}[H]
    \begin{subfigure}{0.5\textwidth}
    \centering
    \includegraphics[width=0.75\textwidth,height=!]{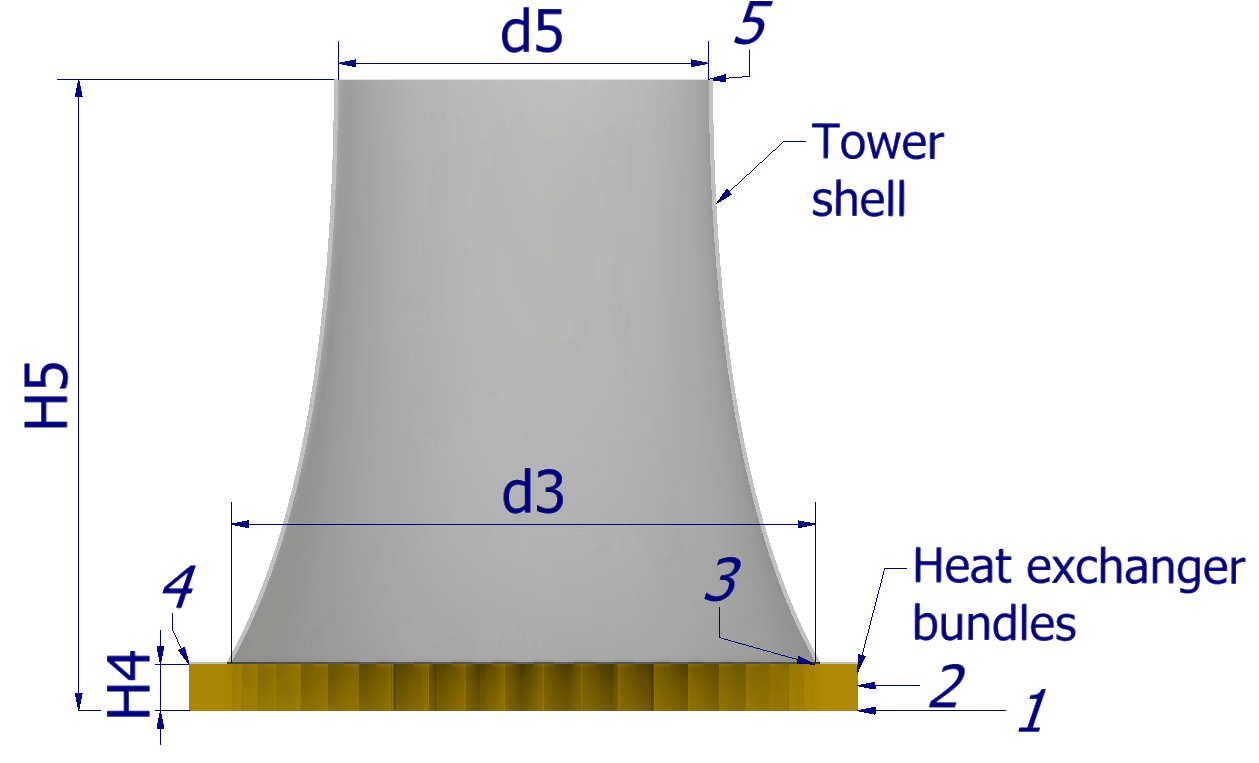}
    \caption{}
    \end{subfigure}
    \begin{subfigure}{0.5\textwidth}
    \centering
    \includegraphics[width=0.255\textwidth,height=!]{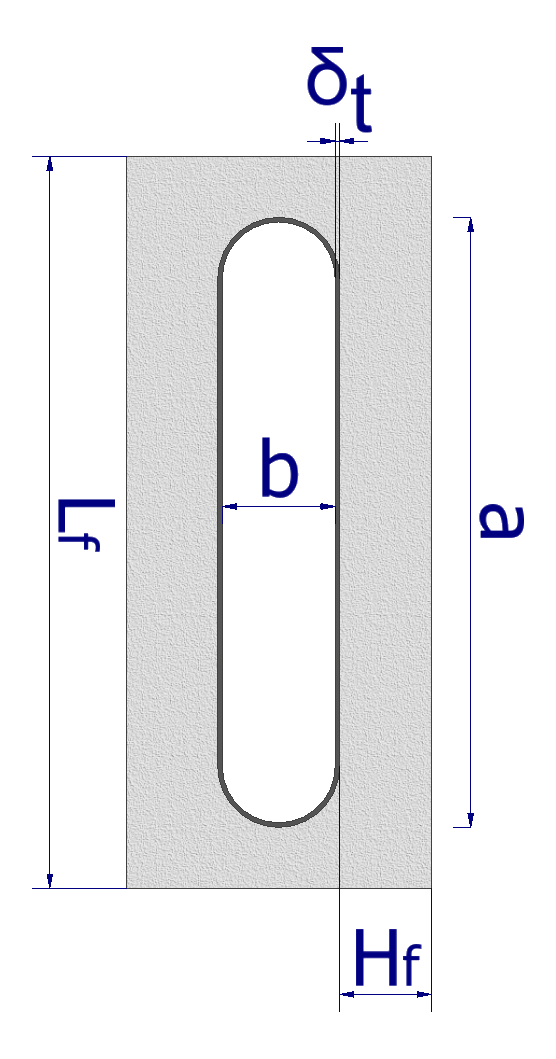}
    \caption{}
    \end{subfigure}
    \caption{Schematic of NDDDCS (a) and effective single-row finned tube (b)}
    \label{fig:tower_tube}
\end{figure}

A schematic of the NDDDCS tower and finned tube geometry is shown in Figure~\ref{fig:tower_tube}, with the reference tower dimensions used in this study listed in Table~\ref{tab:dim}. The tube dimensions, heat transfer- and pressure drop characteristics are identical to those used in previous studies \cite{strydom_HEFAT,strydom_ENFHT}. The total tower height is given by $H_5$, the inlet height is represented by $H_4$ and the inlet- and outlet diameters are indicated by $d_3$ and $d_5$ respectively.    
\vspace{-\baselineskip}

\vspace{-\baselineskip}

\begin{table}[H]
\centering
\small
\caption{NDDDCS reference dimensions at the two considered scales}\label{tab:dim}
\begin{tabular}{l c c c}
\hline
Dimension  &  Symbol& Large-scale & Medium-scale \\
\hline
     Total tower height (m)        &    $H_5$  & 165.0 &  65.0   \\
     Inlet height (m)               &   $H_4$   &  16.5 & 6.5   \\
     Inlet diameter (m)             &  $d_3$  &  165.0 &  65.0   \\
     Outlet diameter (m)              &    $d_5$ &  82.5 & 32.5  \\                       
\hline
\end{tabular}
\end{table}
\vspace{-\baselineskip}

\section*{{\fontsize{12pt}{12pt}\selectfont 3. Mathematical Model}}

The steam- and air-side of the 1-D model are not spatially discretised. The governing equations are solved simultaneously, where required. Transient continuity- and energy equations are derived for the steam-side, while transient continuity-, momentum- and energy equations are derived for the air-side. It is assumed that there is no steam-side pressure drop within the condenser tubes. The density at the next time step on the steam-side is calculated via: 

\vspace{-\baselineskip}
\begin{equation}
  {\rho_{si}}^{t+\Delta t} = \frac{\Delta t}{V_c} \left(\dot{m}_{si} - {\dot{m}_{co}}^{~~{t}} \right) + {\rho_{si}}^t
\end{equation}
\vspace{-\baselineskip}

where ${\rho_{si}}^{t+\Delta t}$ is the steam density at the next time step, $\Delta t$ is the time step, $V_c$ is the total steam-side volume, $\dot{m}_{si}$ the supplied steam mass flow rate boundary condition from the power cycle, ${\dot{m}_{co}}^{~~{t}}$ is the mass flow rate of exiting condensate and ${\rho_{si}}^t$ is the current steam density. The internal energy of the steam is calculated by:

\vspace{-\baselineskip}
\begin{equation}
   {u_{si}}^{t+\Delta t}    = \frac{\Delta t}{{\rho_{si}}^t V_c}\left(\dot{m}_{si} {h_{si}}^t - {\dot{m}_{co}}^{~~{t}} {h_{co}}^t - {\dot{Q}_r}^{~~t} - {u_{si}}^t (\dot{m}_{si} - {\dot{m}_{co}}^{~~{t}} )\right) + {u_{si}}^t
\end{equation}
\vspace{-\baselineskip}

where ${u_{si}}^{t+\Delta t}$ is the internal energy of steam at the next time step, ${u_{si}}^t$ is the current internal energy, ${h_{si}}^t$ and ${h_{co}}^t$ are the respective enthalpies of entering steam and exiting condensate and ${\dot{Q}_r}^{~~t}$ is the steam-side heat transfer rate. The latter is given by:

\vspace{-\baselineskip}
\begin{equation}
    {\dot{Q}_r}^{~~{t}} = \dot{m}_a^{~~t} {c_{pa25}}^t e^t ({T_w}^t - T_{a2}) = {h_c}^t A_c ({T_{si}}^t - {T_w}^t)
    \label{eq:Q_steam_tran}
\end{equation}
\vspace{-\baselineskip}

where ${T_w}^t$ represents the tube wall temperature, $T_{a2}$ the inlet air temperature, ${T_{si}}^t$ represents the saturated inlet steam temperature, $\dot{m}_a^{~~t}$ is the current air mass flow rate, ${c_{pa25}}^t$ is the average specific heat capacity of air, $e^t$ is the heat exchanger effectiveness, ${h_c}^t$ is the condensation heat transfer coefficient and $A_c$ is the total steam-side wetted area. The tube wall temperature is assumed to be uniform between the inner and outer surfaces of the condenser tubes. Rearranging Equation~\ref{eq:Q_steam_tran} to calculate the tube wall temperature yields:

\vspace{-\baselineskip}
\begin{equation}
    {T_w}^t = \frac{{h_c}^t A_c {T_{si}}^t + \dot{m}_a^{~~t} {c_{pa25}}^t e^t T_{a2}}{{h_c}^t A_c + \dot{m}_a^{~~t} {c_{pa25}}^t e^t}
\end{equation}
\vspace{-\baselineskip}

The heat exchanger effectiveness, $e^t$ is given by:

\vspace{-\baselineskip}
\begin{equation}
    e = 1 - \mathrm{exp}\left( \frac{UA^t}{\dot{m}_{a}^{~~t} {c_{pa25}}^t} \right)
\end{equation}
\vspace{-\baselineskip}

Due to the inclusion of the condenser tube wall temperature in the energy balance of Equation~\ref{eq:Q_steam_tran}, the overall heat conductance, $UA$, is equivalent to the air-side heat conductance. The mass flow rate of condensate leaving the condenser tubes is calculated as:

\vspace{-\baselineskip}
\begin{equation}
    \dot{m}_{co}^{~~t} = \dot{Q}_r^{~~t}/{h_{fg}}^t
\end{equation}
\vspace{-\baselineskip}

where ${h_{fg}}^t$ is the latent heat of vaporization. The air-side heat transfer, $\dot{Q}_a^{~~t}$, is given by:

\vspace{-\baselineskip}
\begin{equation}
    \dot{Q}_a^{~~t} = \dot{m}_{a}^{~~t} ({h_{a5}}^t - h_{a2})
\end{equation}
\vspace{-\baselineskip}

where ${h_{a5}}^t$ and $h_{a2}$ are the enthalpies of entering and exiting air, respectively. The air mass flow rate at the next time step is calculated via:

\vspace{-\baselineskip}
\begin{equation}
  \dot{m}_{a}^{~~t+\Delta t} = \frac{A_{35} \Delta t}{L_a} \left(\Delta {P_a}^t - \Delta {P_{L}}^t \right) + \dot{m}_{a}^{~~t}
  \label{eq:momentum_discretized_ch8}
\end{equation}
\vspace{-\baselineskip}

where $\Delta {P_{a}}^t$ is the system driving force, $\Delta {P_{L}}^t$ is the total air-side pressure loss, $A_{35}$ is the average tower flow area and $L_a$ is the air travel distance. The pressure balance in Equation \ref{eq:momentum_discretized_ch8} is equivalent to the draft equation taken from the work of \cite{alma990005992320803436}. Thus, the driving force is:

\vspace{-\baselineskip}
\begin{equation}
 \Delta {P_a}^t =  p_{a1}\left[\left(1-0.00975\frac{H_4}{2T_{a1}} \right)^{3.5}  \left(1-0.00975\frac{H_5 - \frac{H_4}{2}}{{T_{a3}}^t} \right)^{3.5} - \left(1-0.00975\frac{H_5}{T_{a1}} \right)^{3.5} \right] 
\end{equation}
\vspace{-\baselineskip}

where $p_{a1}$ is the ground-level atmospheric pressure, $T_{a1}$ is the corresponding temperature and ${T_{a3}}^t$ is the outlet air temperature after the heat exchangers. The flow resistance is given by:

\vspace{-\baselineskip}
\begin{equation}
\Delta {P_{L}}^t =  \frac{(K_{il} + {K_{he\theta}}^t + {K_{ct}}^t + {K_{ts}}^t)_{he}}{2{\rho_{a25}}^t}
    \left(\frac{\dot{m}_{a}^{~~t}}{A_{fr}} \right)^2
    \left(1-0.00975\frac{H_5-\frac{H_4}{2}}{{T_{a3}}^t}\right)^{3.5} 
    + \frac{({K_{to}}^t + {a_{e5}}^t)}{2{\rho_{a5}}^t}\left(\frac{\dot{m}_{a}^{~~t}}{A_{5}} \right)^2
\end{equation}
\vspace{-\baselineskip}

where ${K_{il}}$ (constant value of 2.5), ${K_{he\theta}}^t$, ${K_{ct}}^t$, ${K_{ts}}^t$, ${K_{to}}^t$ and ${a_{e5}}^t$ represent the non-dimensional inlet louvre-, heat exchanger-, tower-, tower support-, outlet pressure loss and velocity correction factors. The tower outlet air density is represented by ${\rho_{a5}}^t$, the corresponding area is given by $A_5$ and the frontal area of the heat exchangers is represented by $A_{fr}$. The enthalpy of air leaving the NDDDCS is calculated via:

\vspace{-\baselineskip}
\begin{equation}
{h_{a5}}^{t+\Delta t}  =  \frac{\Delta t}{{\rho_{a25}}^t V_a} \left( \dot{m}_{a}^{~~t}  (h_{a2} -  {h_{a5}}^t)  + \dot{Q}_{r}^{~~t}  \right) + {h_{a5}}^{t}
\end{equation}
\vspace{-\baselineskip}

where $V_a$ is half of the internal flow volume. The equations for the loss factors and overall heat conductance are presented in previous papers \cite{strydom_HEFAT,strydom_ENFHT}. 

\section*{{\fontsize{12pt}{12pt}\selectfont 4. Validation and Sensitivity Analysis}}

The 1-D steady state NDDDCS model (SS) was previously validated against the work of Kong et al. \cite{strydom_HEFAT}. Therefore, this study validates the transient 1-D NDDDCS model by comparing the results of the transient model, once it reaches a steady state, to the results of the steady-state model. The large scale tower is used as reference and the same effective boundary conditions are imposed to allow comparison. It is important to note that two transient models were developed: transient steam-side combined with steady-state air-side (TS) and transient steam- and air-side (TT). The SS model was adapted to make use of Equation~\ref{eq:Q_steam_tran} and to neglect the steam-side pressure drop for the sake of consistency between the three models. A constant ambient temperature of 293.15~K (20~$^\circ$C) was selected. The steam mass flow rate (444.98~kg/s) attained by the SS model for a constant saturated steam temperature boundary condition of 323.15~K is assigned as a boundary condition for the TS and TT models. The TT model requires an initial steam temperature and air mass flow rate, specified as the ambient temperature and as effectively zero, respectively. The results are summarized in Table~\ref{tab:NDDDCS_tran_validate}. It is clear from Table~\ref{tab:NDDDCS_tran_validate} that the TS and TT models converge to the same steady-state condition as the SS model. A sensitivity analysis revealed that a time-step of 0.2~s is suitable for this simulation.   

\begin{table}[H]
    \small
    \caption{Summary of steady state results for cross validation of TT model}
    \centering
    \begin{tabular}{l c c c c c c c}
       \hline
       Model   & $T_{si}$~(K) &   $T_{w}$~(K) & $T_{a5}$~(K) & $v_s$~(m/s) & $\dot{Q}$~(MW) &      $\dot{m}_a$~(kg/s) & $\Delta P$~(Pa)  \\
       \hline
        SS    &  323.15      & 322.35       &  315.45     &     99.58        &     1059.92   &   47033.71       &  137.64          \\
        TS    &  323.15      & 322.35       &  315.45     &     99.58        &     1059.92   &   47033.70       &  137.64          \\
        TT    &  323.17      & 322.38       &  315.48     &     99.48        &     1059.90   &   46986.32       &  137.76          \\
    \hline
    \end{tabular}
    \label{tab:NDDDCS_tran_validate}
\end{table}
\vspace{-\baselineskip}


\section*{{\fontsize{12pt}{12pt}\selectfont 5. Results}}

\begin{figure}[H]
\begin{subfigure}{0.5\textwidth}
    \centering
    \includegraphics[width=\textwidth,height=!]{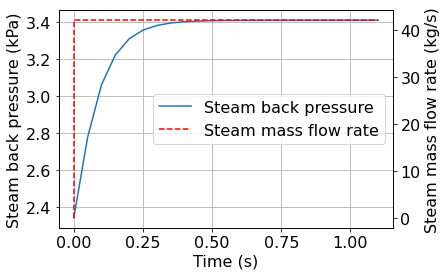}
    \caption{}
    \end{subfigure}
    \begin{subfigure}{0.5\textwidth}
    \centering
    \includegraphics[width=\textwidth,height=!]{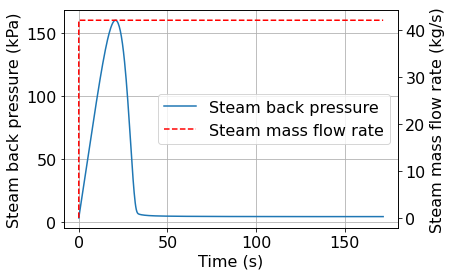}
    \caption{}
    \end{subfigure}
\caption{Large-scale system steam back pressure predicted by TS (a) and TT models (b)}
    \label{fig:ST_TT_comp}
\end{figure}

All properties are calculated using the CoolProp library within the Python environment. The steam mass flow rate ($\dot{m}_{si}$, step input), ambient temperature ($T_{a1}$, 293.15~K) and pressure ($p_{a1}$, 101325~Pa) at ground level are boundary conditions, while the outlet air temperature ($T_{a3}$, 293.15~K), air mass flow rate ($\dot{m}_a$, $\approx$0~kg/s) and saturated steam temperature ($T_{si}$, 295.15~K) are initial conditions. Figure~\ref{fig:ST_TT_comp} serves to illustrate why simulating only the steam-side transiently (TS model) is not adequate to capture the full spectrum of transient start-up effects in NDDDCSs.   

The maximum allowable internal pressure is selected as 160~kPa (a), which is based on an approximate rupture disk burst limit, typically installed on the steam ducting to avoid system overpressure. Results indicate that the large-scale NDDDCS can handle 9.6\% of its maximum steam admission rate as a step input without exceeding this value. Under this admission rate, the TS and TT models calculate vastly different results. Due to the air-side being calculated steadily for the TS model, the airflow through the NDDDCS can respond instantaneously to changes in the steam-side to provide adequate heat rejection. This is not the case for the TT model, as the response time of the air-side is significantly slower than the steam-side. The initial air mass flow rate is not enough to reject the thermal energy provided by the steam, and an acute build up of steam back pressure is observed until the air mass flow rate is sufficient.

\vspace{-\baselineskip}

\begin{figure}[H]
     \begin{subfigure}{0.5\textwidth}
    \centering
    \includegraphics[width=\textwidth,height=!]{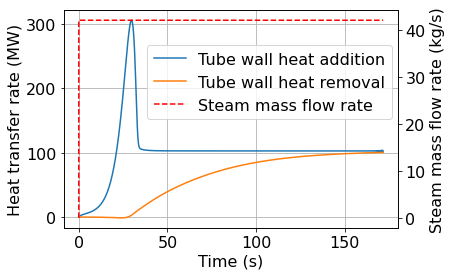}
    \caption{}
    \end{subfigure}
    \begin{subfigure}{0.5\textwidth}
    \centering
    \includegraphics[width=\textwidth,height=!]{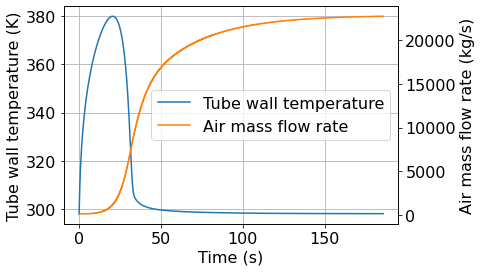}
    \caption{}
    \end{subfigure}
    \caption{Large-scale heat balance (a) and tube wall temperature and air mass flow rate (b) graphs using TT model under maximum steam admission rate step input}
    \label{fig:TT_large_max}
\end{figure}

 Figure~\ref{fig:TT_large_max} (a, b) shows the transient heat balance and tube wall temperature of the large-scale NDDDCS under the same step input previously mentioned.  The heat addition to the tube wall increases rapidly due to a lack of airflow under start-up, leading to a build up in tube wall temperature. An inflection point is reached in the heat addition curve, where the heat removal from the tube wall increases and the heat addition to the tube wall settles as the imbalance in heat transfer starts to reduce, leading to a reduction in steam back pressure. The steam-side settles about five times faster (30~s) than the air-side (150~s).
 
\begin{figure}[H]
    \begin{subfigure}{0.5\textwidth}
    \centering
    \includegraphics[width=\textwidth,height=!]{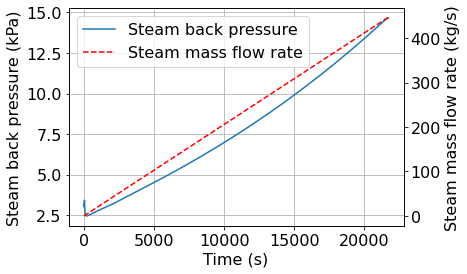}
    \caption{}
    \end{subfigure}
    \begin{subfigure}{0.5\textwidth}
    \centering
    \includegraphics[width=\textwidth,height=!]{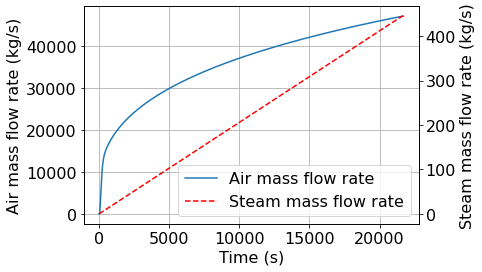}
    \caption{}
    \end{subfigure}
     \begin{subfigure}{0.5\textwidth}
    \centering
    \includegraphics[width=\textwidth,height=!]{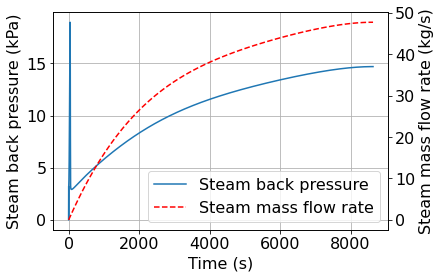}
    \caption{}
    \end{subfigure}
    \begin{subfigure}{0.5\textwidth}
    \centering
    \includegraphics[width=\textwidth,height=!]{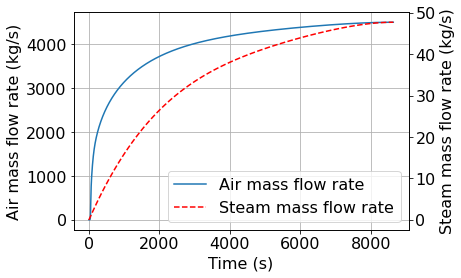}
    \caption{}
    \end{subfigure}
    \caption{Steam back pressure (a, c) and air mass flow rate (b, d) curves under typical start-up load ramp conditions for coal-fired power plant (a, b) and CSP plant (c, d) using TT model}
    \label{fig:TT_large_medium_typ}
\end{figure}

\vspace{-\baselineskip}
The large- (coal-fired) and medium scale (CSP) NDDDCS back pressure and air mass flow rate trends are shown in Figure~\ref{fig:TT_large_medium_typ} for typical start-up steam admission rates, acquired from proprietary sources.

The large- and medium-scale NDDDCSs experience controlled spikes in steam back pressure (Figure~\ref{fig:TT_large_medium_typ} (a, c)) initially due to airflow in the tower not being established yet, after which the air mass flow rate increases rapidly, as seen in Figure~\ref{fig:TT_large_medium_typ} (b, d). As the air mass flow rate continues to increase, the NDDDCSs can easily supply the required heat rejection rate to match the transient cycle requirements, which controls the back pressure. This indicates that for a typical coal-fired or CSP plant, the NDDDCS will not impose transient operation limits, as the fastest cycle load ramp rates are still slower than the response time of the cooling system. This belays past concerns about the transient response time of natural draft steam condensing systems, which rely on the interplay between the condensing process and draft creation. 

\section*{{\fontsize{12pt}{12pt}\selectfont 6. Conclusion}}

This study investigated the transient start-up characteristics of NDDDCSs at large- and medium-scales under transient start-up and load ramp scenarios, for windless atmospheric conditions. Results indicate that the steam-side responds much faster than the air-side of the system during unsteady conditions. The study finds that for the typical transient scenarios experienced at a coal-fired or CSP plant, the NDDDCS is able to respond rapidly enough to avoid back pressure increases that would limit turbine output during the load ramp. 

\section*{{\fontsize{12pt}{12pt}\selectfont Acknowledgements}}
This work is supported by the National Research Foundation (NRF) of South Africa and the Solar Thermal Energy Research Group (STERG) of the University of Stellenbosch. 

\section*{{\fontsize{12pt}{12pt}\selectfont Nomenclature}}
\vspace{-\baselineskip}

\begin{multicols}{2}
\begin{spacing}{0}
\small
\nomenclature[P00]{\textbf{Parameters}}{}
\nomenclature[P01]{$A$}{Area ({$\mathrm{m^2}$})}
\nomenclature[P02]{$a$}{Major axis or velocity correction factor ({$\mathrm{m~or}~-$})}
\nomenclature[P03]{$b$}{Minor axis ({$\mathrm{m}$})}
\nomenclature[P04]{$c$}{Specific heat capacity ({$\mathrm{J/kg\cdot K}$})}
\nomenclature[P05]{$d$}{Diameter ({$\mathrm{m}$})}
\nomenclature[P07]{$g$}{Gravitational acceleration ({$\mathrm{m/s^{2}}$})}
\nomenclature[P08]{$H$}{{Height ($\mathrm{m}$})}
\nomenclature[P09]{$h$}{ Heat transfer coefficient or latent heat of vaporization ({$\mathrm{W/m^{2}\cdot K}$} or {$\mathrm{J/kg}$})}
\nomenclature[P11]{$K$}{Loss factor ({$\mathrm{-}$})}
\nomenclature[P12]{$k$}{Thermal conductivity ({$\mathrm{W/m\cdot K}$)}}
\nomenclature[P13]{$L$}{Length ({$\mathrm{m}$})}
\nomenclature[P15]{$\dot{m}$}{Mass flow rate ({$\mathrm{kg/s}$})}
\nomenclature[P18]{$p$}{Pressure or pitch ({$\mathrm{Pa~or~m}$})}
\nomenclature[P19]{$Pr$}{Prandtl number ({$\mathrm{-}$})}
\nomenclature[P20]{$\dot{Q}$}{Heat transfer rate ({$\mathrm{W}$})}
\nomenclature[P24]{$T$}{Temperature ({$\mathrm{K}$})}
\nomenclature[P25]{$UA$}{Overall heat conductance ({$\mathrm{W/K}$})}
\nomenclature[P28]{$v$}{Velocity ({$\mathrm{m/s}$})}

\nomenclature[G0]{\textbf{Greek Symbols}}{}
\nomenclature[G1]{$\Delta$}{Change or thickness ({$\mathrm{-~or~m}$})}
\nomenclature[G2]{$\delta$}{Thickness ({$\mathrm{m}$})}
\nomenclature[G3]{$\epsilon$}{NTU-effectiveness ({$\mathrm{-}$})}
\nomenclature[G4]{$\mu$}{Dynamic viscosity ({$\mathrm{kg/m\cdot s}$})}
\nomenclature[G5]{$\rho$}{Density ({$\mathrm{kg/m^3}$})}

\nomenclature[S0]{\textbf{Subscripts}}{}
\nomenclature[S1]{$a$}{Air, Atmospheric}
\nomenclature[S2]{$c$}{Condensate, Cooling}
\nomenclature[S3]{$f$}{Fin}
\nomenclature[S4]{$fr$}{Frontal}
\nomenclature[S5]{$he$}{Heat exchanger}
\nomenclature[S6]{$i$}{Inlet, In}
\nomenclature[S7]{$l$}{Louvre}
\nomenclature[S8]{$L$}{Loss}
\nomenclature[S9]{$o$}{Outlet, Out}
\nomenclature[S10]{$r$}{Radiator}
\nomenclature[S11]{$s$}{Steam, Support}
\nomenclature[S12]{$t$}{Tube, Tower}
\nomenclature[S13]{$w$}{Wall}
\nomenclature[S14]{$3$}{Tower inlet diameter}
\nomenclature[S15]{$4$}{Tower inlet height}
\nomenclature[S16]{$5$}{Tower outlet diameter}

\nomenclature[X0]{\textbf{Superscripts}}{}
\nomenclature[X1]{$t$}{Time}
\nomenclature[X2]{$\Delta$}{Change}

\nomenclature[Z0]{\textbf{Abbreviations}}{}
\nomenclature[Z2]{ACC}{~~~~~Air-cooled condenser}
\nomenclature[Z4]{NDDDCS}{Natural draft direct dry cooling system}
\nomenclature[Z1]{3-D}{~~~~~3-dimensional}
\nomenclature[Z3]{CFD}{~~~~~Computational fluid dynamics}
\printnomenclature
\end{spacing}
\end{multicols}

\vspace{-\baselineskip}
\bibliography{ref}
\end{document}